\documentclass[letterpaper,twocolumn,10pt]{article}

\usepackage{usenix2019_v3}

\usepackage{amsmath}
\usepackage{graphicx}
\usepackage{svg}

\usepackage{filecontents}

\begin{document}

\date{}

\title{\Large \bf TonY: An Orchestrator for Distributed Machine Learning Jobs}

\author{
{\rm Anthony Hsu, Keqiu Hu, Jonathan Hung, Arun Suresh, Zhe Zhang}\\
{\rm \normalsize LinkedIn} \\
{\rm \normalsize \{ahsu,khu,jhung,asuresh,zezhang\}@linkedin.com}
} 

\maketitle

\let\thefootnote\relax\footnotetext{To appear in OpML '19}

\begin{abstract}
Training machine learning (ML) models on large datasets requires considerable computing power.
To speed up training, it is typical to distribute training across several machines,
often with specialized hardware like GPUs or TPUs. Managing a distributed training job is complex
and requires dealing with resource contention, distributed configurations, monitoring, and fault tolerance.
In this paper, we describe TonY, an open-source orchestrator for distributed ML jobs
built at LinkedIn to address these challenges.
\end{abstract}

\section{Introduction}

The past couple of decades have seen an explosion in "Big Data" systems for storing and processing
data. Some widely used systems include MapReduce~\cite{dean2008mapreduce}, Hadoop Distributed File System~\cite{shvachko2010hadoop},
and Spark~\cite{zaharia2012resilient}.
The scale of these systems has made it possible to store petabytes of data and do large-scale
ML.

Many features on the LinkedIn website are powered by ML, including People You May Know,
Job Recommendations, the News Feed, and Learning Recommendations. Many of these features are powered
by ML techniques such as boosted decision trees~\cite{chen2016xgboost} and generalized linear models~\cite{zhang2016glmix}.

To boost the accuracy of predictions, ML engineers have started experimenting with non-linear models
such as neural networks~\cite{Goodfellow-et-al-2016} to capture more complex relationships in the data.
Programming these neural networks in a generic language is tedious and error-prone. To address this,
many frameworks have been created to simplify the construction and training of neural
networks. These frameworks include DistBelief~\cite{dean2012large} 
and its successor TensorFlow~\cite{abadi2016tensorflow}, Theano ~\cite{bergstra2010theano}, Caffe~\cite{jia2014caffe}, PyTorch~\cite{paszke2017automatic}, and Keras~\cite{chollet2015keras}.

An ML engineer will often begin model development by developing on a single machine. One popular tool is
a "notebook" program such as Jupyter~\cite{Kluyver:2016aa} or Zeppelin~\cite{Zeppelin} that allows an ML engineer
to interactively explore the data and test out fragments of their models. This works when experimenting
on a sample of the data. However, to validate a new model, they generally need to train and test their
model on the full dataset, which may be petabytes in size and would take too long to train on a single machine.
To scale up their training, they need to divide the data across multiple machines and
train in parallel~\cite{ben2018demystifying}.

Most ML frameworks provide APIs for doing distributed training. However, to make use of multiple machines, an ML engineer still
has to copy their program to each host, set the appropriate environment variables and configurations for
distributed training on each host, and then launch their training program on each host. This ad-hoc process faces several challenges:

\begin{itemize}
\item \textbf{Resource contention.} ML engineers sharing the same pool of unmanaged machines
fight for the same memory, CPU, and GPU resources. Consequently, jobs may fail with out-of-memory
exceptions or errors allocating GPUs.
\item \textbf{Tedious and error-prone configuration.} Setting up a distributed training job
requires copying configurations to all hosts and it is hard to verify and update these
configurations.
\item \textbf{Lack of monitoring.} While the job is running, it is difficult to monitor
its global progress.
\item \textbf{Lack of fault tolerance.} Transient errors are hard to debug and require
manual restarts.
\end{itemize}

To address these challenges, we built and open-sourced TonY~\cite{hung2018tony}, an orchestrator that interacts with a cluster scheduler
to launch and manage distributed training jobs.

\section{Architecture}

\begin{figure}[htbp]
  \centering
  \def\svgwidth{\columnwidth}
  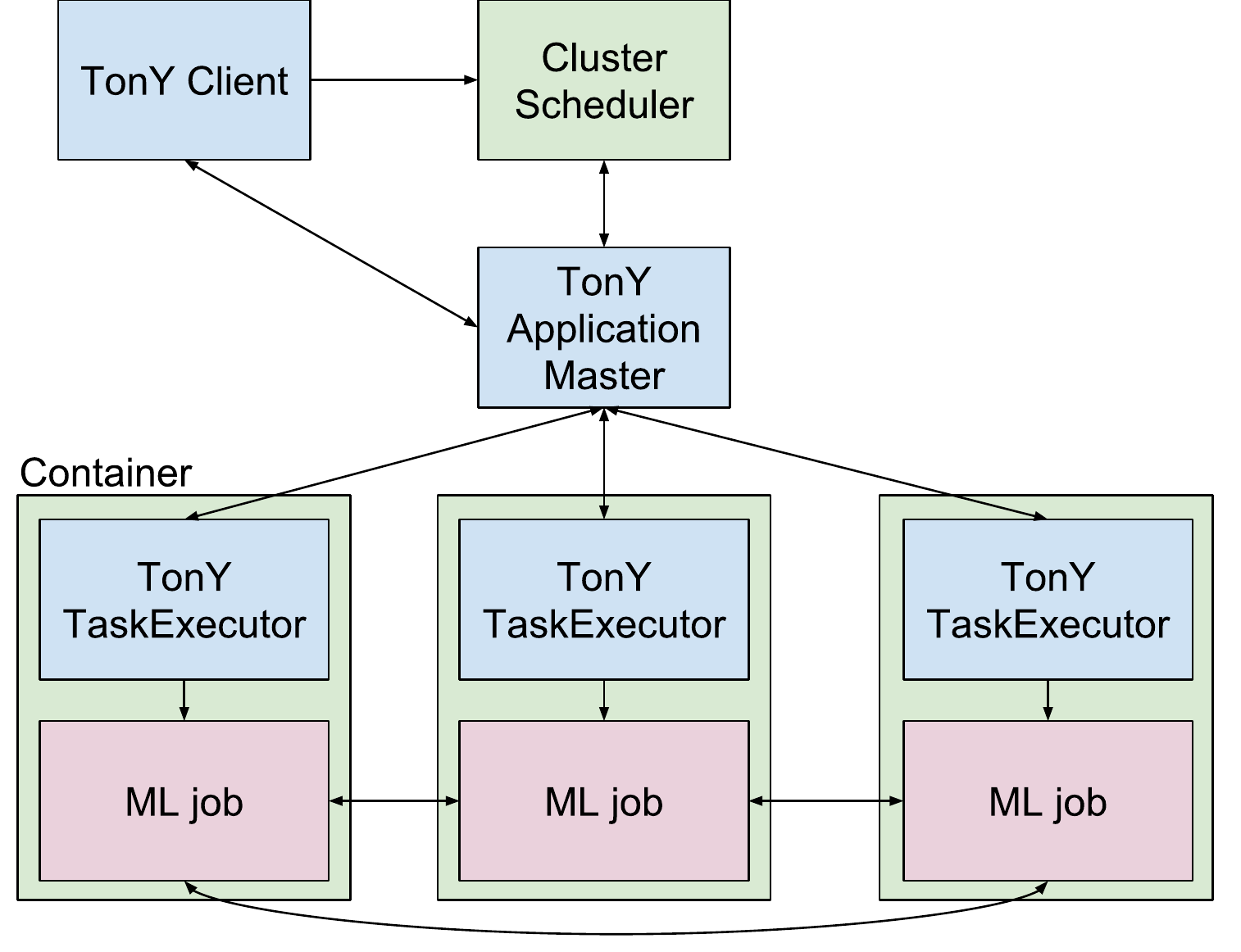
  \caption{\label{fig:tony-architecture} TonY's architecture.}
\end{figure}

TonY consists of a client for submitting jobs to a scheduler and an application
that runs in the scheduler. Users use the client to submit their ML
jobs, and the application handles allocating resources, setting up configurations,
and launching the ML job in a distributed fashion. The client interface is
generic and its implementation can support submitting to multiple schedulers.
The scheduler implementation can be changed without requiring users to update their
ML or client submission code.

For our initial implementation of TonY, we added support for running distributed
\underline{T}ensorFlow jobs \underline{on} Hadoop \underline{Y}ARN
(Yet Another Resource Negotiator)~\cite{vavilapalli2013apache} (hence the name TonY),
as these were the most commonly used ML framework and scheduler, respectively,
at LinkedIn.

The overall architecture of TonY is presented in Figure~\ref{fig:tony-architecture}.
We present the client and cluster components of TonY in more detail in the following
subsections.

\subsection{TonY Client}

The TonY client is the library users use to launch their distributed ML jobs.
Users describe in an XML file the resources required by their
job. For TensorFlow, this might include the number of worker and parameter
server instances as well as how much memory and how many GPUs per
instance. If needed, users can also specify additional configurations for
the underlying scheduler. In the case of YARN, this might include specifying
the queue~\cite{orgqueue2016} or node label~\cite{karanasos2018yarnadvances} (e.g.: high-memory)
to run on.

Users will also provide the path to their ML program as well as the virtual environment
or Docker image~\cite{merkel2014docker} in which their program should run on the cluster. Additionally,
users can specify properties such as model-specific hyperparameters,
input data, and output location via command-line arguments passed to the TonY client.

Often, distributed ML jobs will be run as part of a larger workflow
that includes data preprocessing and model deployment. To simplify integration
into existing workflows, we built a TonY plugin for one such
workflow manager, Azkaban~\cite{Azkaban}, that lets users add distributed ML jobs
in the same workflow alongside Spark, MapReduce, and other jobs.

\subsection{TonY Cluster Application}

When the user runs the TonY Client to submit their job, the client will package
the user configurations, ML program, and virtual environment into an archive
file that it submits to the cluster scheduler.

The TonY Client will launch a master program in the cluster scheduler. In our
initial implementation supporting Hadoop's YARN scheduler, we launch a TonY
ApplicationMaster (AM) in a YARN container. The AM then negotiates
with YARN's ResourceManager (RM) to request all the other containers (e.g.: worker
and parameter server tasks) needed by the ML job. The AM handles heterogeneous
resource requests for different task types, such as requesting containers with
GPUs for the worker tasks but requesting CPU-only containers for the parameter
server tasks.

Once the task containers are allocated by the RM to the TonY AM, it then launches
a TaskExecutor in each task container. This TaskExecutor will allocate a port
for its task to run on and register this port with the AM. Upon receiving
registration from all TaskExecutors, the AM will construct a global cluster spec that it
will then send back to every TaskExecutor. Each TaskExecutor will then
set the global cluster spec along with task-specific configuration in environment variables
before spawning the ML job as a child process. Once all the ML jobs start up,
they will communicate and coordinate with one another via the ML framework's distributed
protocol (whether that be RPC, MPI, etc.), and the TaskExecutors will
monitor the task processes and heartbeat back to the AM. When the task processes finish,
the TaskExecutor will register the final exit status with the AM before terminating.

The TaskExecutor for the first worker task will also allocate a port for launching a visualization
user interface such as TensorBoard for monitoring the running job. This also gets
registered with TonY AM. This user interface URL, along with links to all the other task logs,
is sent back to the TonY Client so that users can directly access the visualization UI and
task logs from one place.

Finally, if any task fails, the TonY AM will automatically tear down the remaining
tasks, request new task containers, setup a new global cluster spec, and relaunch the tasks.
The ML tasks can then restore from the last checkpoint and continue training.

\section{Discussion}

Previously, ML engineers had to write ad-hoc scripts to launch distributed ML jobs
on a pool of machines, with no resource guarantees or isolation from other users.
Now, using TonY, users can configure their job once and rely on TonY to negotiate with a cluster scheduler for guaranteed resources.

The TonY master handles all the distributed setup and provides a central place to monitor
and visualize the training job. It also ensures fault tolerance by restarting distributed jobs
in case of transient task failures.

The master and TaskExecutor orchestration framework is also an ideal place to instrument
the ML tasks and collect metrics about the tasks' performance and resource utilization.
These statistics could be aggregated and analyzed in a UI such as Dr. Elephant~\cite{rai2016drelephant}
to suggest new settings for the ML jobs that would improve performance and resource utilization.
We are currently implementing these new features in TonY and plan to discuss them more in future work.

\bibliographystyle{plain}
\bibliography{\jobname}

\end{document}